\documentclass[12pt,preprint,psfig]{aastex}

\shorttitle{GRBs and Cosmology}
\shortauthors{Petrosian, Bouvier \& Ryde}

\begin{document}


\title{Gamma-ray Bursts as Cosmological Tools}


\author{Vah\'{e} Petrosian\altaffilmark{1,2}, Aur\'{e}lien Bouvier
\altaffilmark{2}
and Felix Ryde \altaffilmark{3}}
\affil{Center for Space Science and Astrophysics, Department of Physics,
Stanford
University, Stanford, CA 94305}


\altaffiltext{1}{Department of Applied Physics, Stanford University, Stanford,
CA,
94305 email: vahe@astronomy.stanford.edu}
\altaffiltext{2}{KIPAC, Stanford Linear Accelerator Center}
\altaffiltext{3}{Stockholm Observatory, Stockholm, Sweden}


\newcommand{\myemail}{vahe@astronomy.stanford.edu}

\newcommand\ie{{\it i.e.\,}}
\newcommand\ea{{\it et al.\,}}
\newcommand\eg{{\it e.g.\,}}
\newcommand\D{\Delta}
\newcommand\g{\gamma}
\newcommand\rel{relativistic \,}
\newcommand\be{\begin{equation}}
\newcommand\ee{\end{equation}}
\newcommand\cl{\centerline}
\newcommand\bs{\bigskip}
\newcommand\ms{\medskip}
\newcommand\np{\newpage}
\newcommand{\bea}{\begin{eqnarray}}
\newcommand{\eea}{\end{eqnarray}}
\newcommand{\eiso}{{\cal E}_{\rm iso}}
\newcommand{\egamma}{{\cal E}_\gamma}
\newcommand{\ep}{E_p}
\newcommand{\epo}{E_p^{\rm obs}}
\newcommand{\tjet}{\theta_{\rm jet}}

\baselineskip=13pt

\begin{abstract}

In recent years there has been considerable activity in using gamma-ray bursts
as cosmological probes for determining global cosmological parameters
complementing results from type Ia supernovae and other methods. This requires a
characteristics of the source to be a standard candle. We show that contrary to
earlier indications the accumulated data speak  against this possibility. Another
method would be to use correlation between a distance dependent and a  distance
independent variable to measure distance and determine cosmological parameters
as is done using Cepheid variables and to some extent Type Ia supernovae. Many
papers have dealt with the use of so called Amati relation, first predicted by
Lloyd, Petrosian and Mallozzi, or the Ghirlanda relation for this purpose. We
have argued that these procedure involve many unjustified assumptions which if
not true could invalidate the results. In particular, we point out that many
evolutionary effects can affect the final outcome. In particular, we demonstrate
that the existing data from {\it Swift} and other earlier satellites show that
the gamma-ray burst may have undergone luminosity evolution. Similar evolution
may be present for other variables such as the peak photon energy of the total
radiated energy. Another out come of our analysis is determination of the
luminosity function and the comoving rate evolution of gamma-ray bursts which
does not seem to agree with the cosmic star formation rate. We caution however,
that the above result are preliminary and includes primarily the effect of
detection threshold. Other selection effects, perhaps less important than this,
are also known to be present and must be accounted for. We intend to address
these issues in future publications.

\end{abstract}

\section{INTRODUCTION}

The change in our understanding of gamma-ray bursts (GRBs) in less than a decade
has been
unprecedented. We have gone from groping for ways to determine their distances
(from solar system to cosmological scales) to attempts to use them  as
cosmological probes. Observations by instruments on board a series of satellites
starting
with {\it BeppoSAX} and continuing with {\it HETE}, {\it INTEGRAL} and
{\it Swift}, have been the primary source of this change. The higher spatial
resolution of these instruments has allowed the measurement of redshifts of many
well-localized GRBs, which has in turn led to several
attempts to discover some emission characteristics  which appears to be
a ``standard candle" ({\bf SC} for short), or shows a well defined
correlation (with a small dispersion) with another distance independent
measurable characteristic. One can use such relations to determine the distances
to
GRBs in a manner analogous to the use of the Cepheid variables.
Example of this are the lag-luminosity and variability-luminosity
relations (Norris et al 2000, Norris 2002, Fenimore \&  Ramirez-Ruiz
2000, Reichart et al 2001) which were exploited for determining some
cosmological aspects of these sources (Lloyd, Fryer \& Ramirez-Ruiz 2002,
Kocevski \& Liang 2006) using the methods developed by
Efron \& Petrosian (1992, 1994, 1999). More recently there has
been a flurry of activity dealing with the observed relation between the
peak energy $\ep$ of the $\nu F_\nu$ spectrum and the total (isotropic)
gamma-ray  energy output
$\eiso$ ($\ep\propto\eiso^{\eta}$, $\eta \sim 0.6$) predicted by Lloyd \ea
(2000, {\bf LPM00}) and established to be the case by Amati \ea (2002) (see also
Lamb \ea 2004, Attiea et al 2004). Similarly Ghirlanda \ea (2004a) has shown a
correlation with smaller dispersion between $\ep$ and the beaming corrected
energy $\egamma$ ($E_p \propto \egamma^{\eta'}$, $\eta' \sim 0.7$) where:

\be
\egamma=\eiso \times \frac{(1-cos(\tjet))}{2}
\ee

and $\tjet$  is the half width of the jet. 

These are followed by many attempt to use these relations
for determining  cosmological parameters (Dai \ea 2004,Ghirlanda \ea 2004b, 2005
[Gea05],  Friedman \& Bloom 2005, [FB05]). There are, however, many
uncertainties associated with the claimed relations and even more with
the suggested cosmological tests. The purpose of this paper is to investigate
the utility of GRBs as cosmological tools either as SCs or via some correlation.
In the next section we review the past and current status of the first
possibility
and in \S3 we discuss the energy-spectrum correlation and whether it can be used
for  cosmological model parameter determination. Finally in \S4, we address the
question of cosmological luminosity and rate density evolution of GRB based on
the existing sample with known redshifts.

\section{Standard Candle?}

The simplest method of determining cosmological parameters is through SCs. Type
Ia supernovae (SNIa) are a good example of this. But currently their
observations
are limited to relatively nearby universe (redshift $z<2$). Galaxies and active
galactic nuclei (or AGNs), on the other hand, can be seen to much higher
redshifts ($z>6$) but are not good SCs. GRBs are observed to similar redshifts
and can be detected to even higher redshifts by current instruments. So that if
there were SCs they can complement the SNIa results. In general GRBs show
considerable dispersion in their intrinsic characteristics. The first indication
that GRBs might be SCs came from Frail \ea (2001) observation showing that for a
sample of 17 GRBs  the
dispersion of the distribution of $\egamma$ is significantly smaller than that
of $\eiso$.
The determination of $\egamma$ requires a well defined light curve with a
distinct steepening. The jet opening $\theta_{jet}$ depends primarily on the
time of the steepening and the bulk Lorentz factor, but its exact value is model
dependent and depends also weakly on $\eiso$ and the density of the background
medium

\be
\label{theta}
\theta_{jet}=0.101 \mbox{ radian} \times \left( \frac{t_{break}}{1 \mbox{ day}}
\right)^{3/8} \left( \frac{\eta}{0.2} \right)^{1/8} \left( \frac{n}{10 \mbox{
cm}^{-3}} \right)^{1/8} \left( \frac{1+z}{2} \right)^{-3/8} \left( \frac{{\cal
E}_{iso}}{10^{53} \mbox{ ergs}} \right)^{-1/8},
\ee

\begin{figure}[htbp]
\begin{center}
\includegraphics[width=18cm]{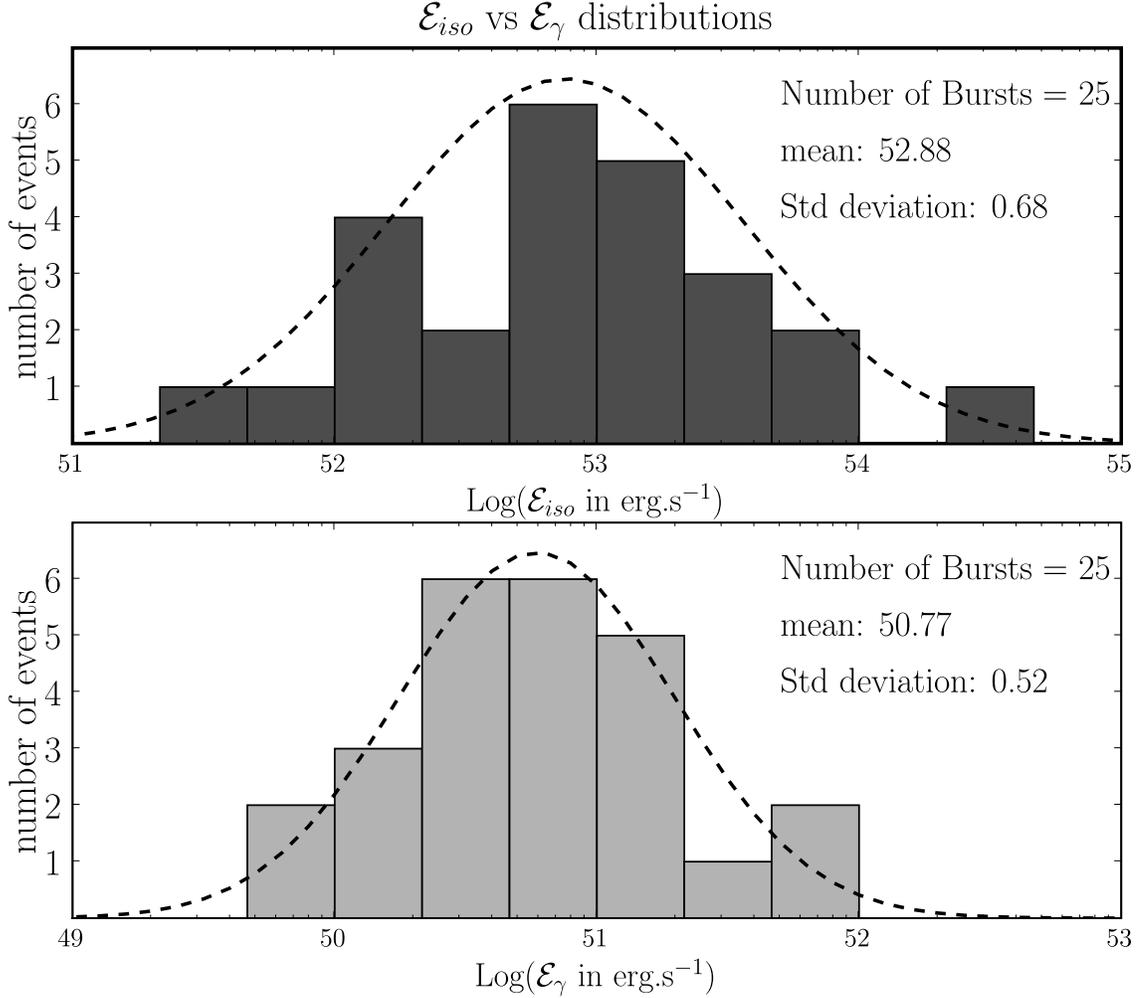}
\end{center}
\caption{Distribution of $\eiso$ and $\egamma$ for 25 pre-Swift GRBs with
evidence for a jet-break and beaming. $\egamma$ distribution is shifted by about
two orders of magnitudes compared to $\eiso$ distribution due to the beaming
factor correction. However the dispersion of the two distribution are very
similar ($\sigma_{\eiso} =0.68$, $\sigma_{\egamma} =0.52$) and broad indicating
that GRBs cannot be assumed to be {\bf SCs}}
\label{1}
\end{figure}

In Figure \ref{1} we show the distribution of $\eiso$ and $\egamma$ for  25
pre-{\it Swift} GRBs (mostly compiled in FB05). As evident, there is little
difference between the two distributions (except for their mean values) and
neither characteristics is anywhere close to being a SC.

We have calculated the jet angle $\tjet$ (from equation \ref{theta}) for 25
pre-{\it Swift} GRBs with relatively well defined $t_{break}$. Assuming a
gamma-ray efficiency $\eta = 0.2$ and a value of circum burst density estimated
from broadband modeling of the lightcurve when available (otherwise we use the
default value of $n=10$ cm$^{-3}$).

Unfortunately fewer than expected {\it Swift} GRBs  have optical light curves
and their X-ray light curves show considerable structure (several breaks and
flaring activity)
with several GRBs showing no sign of jet-break or beaming (Nousek \ea 2006).
This has brought the whole idea of jet breaks and calculating $\egamma$ into
question.
The upshot of this is that {\it Swift} $\egamma$, like $\eiso$ also has a broad
distribution extending
over two decades. Thus,  any cosmological use of GRBs must
include the effects of the breadth of the distributions%
\footnote{It should be also noted that there is an observational
bias in favor of detecting smaller jet angles (\ie earlier jet-breaks), so that
the population as a whole (including those with very late jet-breaks)
will have even a broader distribution. Note also that a {\bf SC} $\egamma$ means
that the Ghirlanda relation would have an index $\eta' \sim 0$}.

\section{Correlations}

When addressing the correlation between any two variables, one should
distinguish between a
one-to-one relation and a statistical correlation. In general the correlation
between two variables (say, $\ep$ and $\eiso$) can be described by a
bi-variate distribution  $\psi(\ep, \eiso)$. If this is a
separable function, $\psi(\ep, \eiso) = \phi(\eiso)\zeta(\ep)$, then the two
variables are
said to
be uncorrelated. A correlation is present if some characteristic (say the mean
value)
of one variable depends on the other:
\eg $\langle\ep\rangle=g(\eiso)$. Only in the absence of dispersion there will
be
a one-to-one relation;
$\zeta(\ep)=\delta(\ep-g(\eiso))$.

In general, the determination  of the exact
nature of the correlation is complicated by the fact that the extant data
suffers from many observational selection biases and truncations. An obvious
bias is that most sample are limited to GRBs with peak fluxes above some
threshold. There are also biases in the
determination of $\epo$ (see \eg Lloyd \& Petrosian 1999; {\bf LP99}). The
methods devised by Efron \& Petrosian (1992, 1994) are particularly suitable for
determination of correlations in such complexly truncated data.

The first indication of a correlation between
the energetics and spectrum of
GRBs came from Mallozzi et al (1995), who reported a
correlation between observed peak flux $f_p$ and $\epo$.
A more comprehensive analysis  by {\bf LPM00}, using the above mentioned
methods, showed that a
similar correlation also exist  between the observed total energy fluence
$F_{\rm tot}$ and $\epo$. Both these quantities depend on the redshift $z\equiv
Z-1$;
\be
\label{eiso}
F_{\rm tot}=\eiso/(4\pi d_m^2 Z), \,\,\,\,\,\, {\rm and} \,\,\,\,\, \epo=\ep/Z,
\ee
Here $d_m$ is the metric distance, and for a flat universe
\be
\label{dm}
d_m(Z) = (c/H_0)\int^Z_1 dZ'(\Omega (Z'))^{-1/2}, \,\,\,\,\, {\rm with}
\,\,\,\,\, \Omega (Z) = \rho(Z)/\rho_0,
\ee
describing the evolution of the
total energy density $\rho(z)$  of
all substance (visible and dark matter, radiation, dark energy or the
cosmological
constant).
{\bf LPM00} also showed that the correlation expected from these
interrelationships is not sufficient to account for the observed correlation,
and  that there must be an
intrinsic correlation between $\ep$ and $\eiso$. Without knowledge of
redshifts LPM00 predicted the relation ${\cal E}_{iso} \propto E_p^{0.5}$ which
is very close
0.5, which is very similar
to the so-called Amati relation obtained for GRBs with known
redshifts. However, it should be emphasized that the LPM00 result implies a
statistical correlation  and
not a one-to-one relation needed for using GRBs as a reliable distance
indicator.
Nakar \& Piran (2004, 2005) and Band \& Preece
(2005) have shown convincingly that  the claimed tight one-to-one
relations cannot be valid for all GRBs. We believe that the
small dispersion seen in GRBs with known redshifts is due to
selection effects arising in the localization
and redshift determination processes: {\it e.g.}, these GRBs
may represent the upper envelope of the
distribution. A recent analysis  by Ghirlanda \ea
(2005) using
pseudo-redshift  shows a much broader dispersion (as in LPM00).
The claimed  tighter Ghirlanda relation,
could be due to additional correlation between the jet opening angle $\tjet$
and $\eiso$, $\ep$, or both. However, as mentioned above the picture of jet
break and measurements of $\theta_{jet}$ and ${\cal E}_\gamma$ is a confusing
state in view of {\it Swift} observation.

\begin{figure}[htbp]
\begin{center}
\includegraphics[width=18cm]{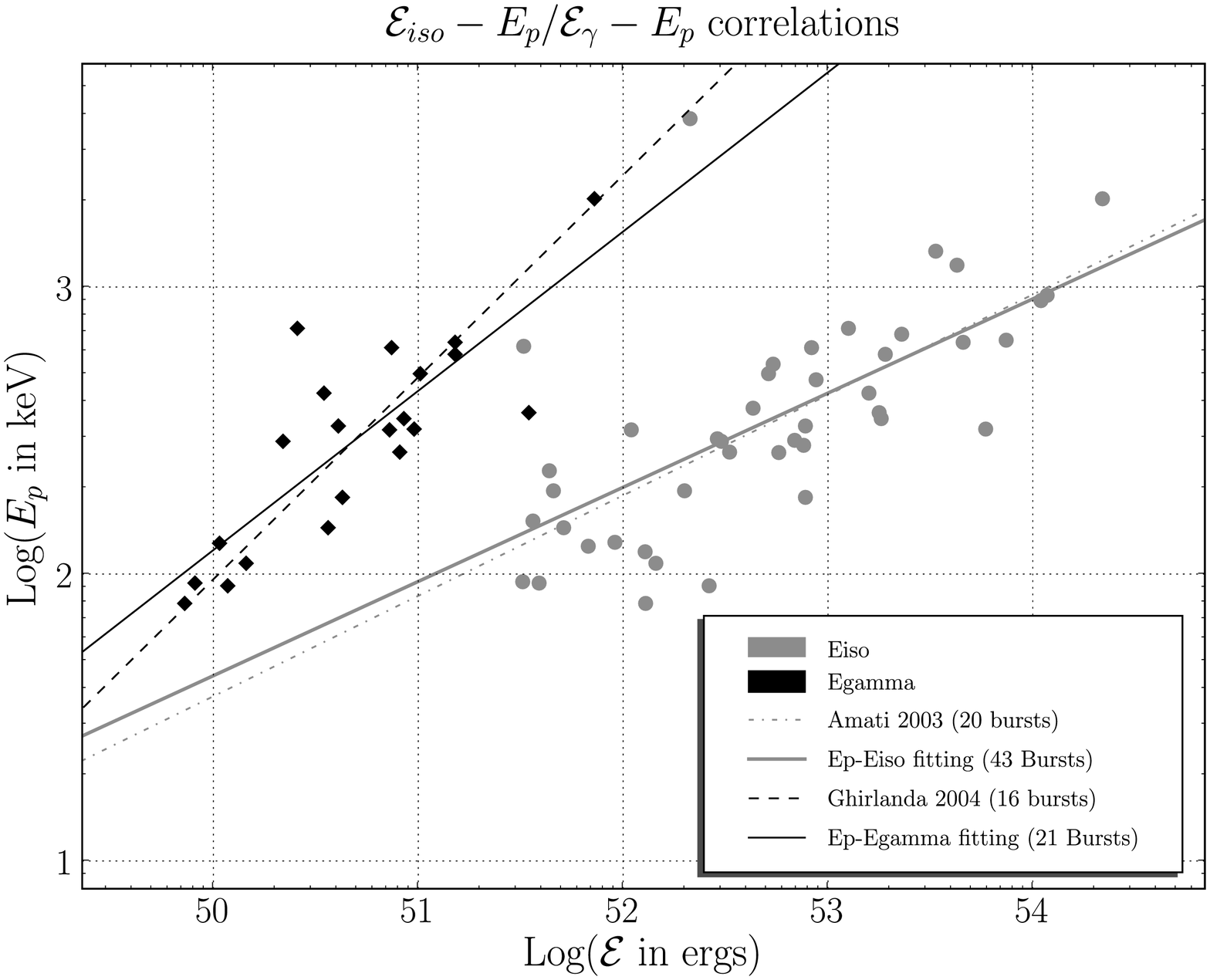}
\end{center}
\caption{$E_p-{\cal E}_{iso}$ and $E_p-{\cal E}_{\gamma}$ correlations. The 43
gray circles are all the bursts from our sample that had good enough  spectral
observations to find the energy peak of the $\nu F_{\nu}$ spectrum, and the 21
black diamonds are a subset of those bursts with a jet break found in their
optical lightcurve. Solid lines are the best fit we find for the two
correlations ($E_p \propto {\cal E}_{iso}^{\eta}$, $\eta \sim 0.328 \pm 0.036$
and $E_p \propto \egamma^{\eta'}$, $\eta' \sim 0.555 \pm 0.089$) and dashed
lines are the best fit found by Amati \ea 2003 with 20 bursts ($E_p \propto
{\cal E}_{iso}^{0.35}$) and Ghirlanda \ea 2004 with 16 bursts ($E_p \propto
{\cal E}_{\gamma}^{0.70}$).}
\label{2}
\end{figure}

We have reanalyzed the existing data and determined the parameters of the Amati
and Ghirlanda relations. In Figure \ref{2} we show the the $\eiso$ (and
$\egamma$ excluding some outliers) vs $\ep$ for all GRBs with known redshifts
(and $\tjet$). We compute best power law fit for both these correlations and we
describe the dispersion around it by the standard deviation. For the $E_p -
{\cal E}_{iso}$ we find: $E_p \propto {\cal E}_{iso}^{\eta}$, $\eta \sim 0.328
\pm 0.036$ and $\sigma_{iso} = 0.286$. For $E_p - \egamma$ correlation we find:
$E_p \propto \egamma^{\eta'}$, $\eta' \sim 0.555 \pm 0.089$ and $\sigma_{\gamma}
= 0.209$. We find that additional data has reduced the significance of the
correlations or has increased the dispersions (compare our values of
$\sigma_{iso} \sim .... $ and $\sigma_{\gamma} \sim .... $ obtained by Amati \ea
and Ghirlanda \ea). This is contrary to what one would expect for a sample with
a true correlation.
From this we conclude that, as predicted by LPM00, there is a strong correlation
between $\ep$ and $\eiso$ (or $\egamma$), but for the population  GRBs
as a whole both variables have a broad distribution and most
GRBs do not obey the tight relations claimed earlier.

\subsection{Correlations and Cosmology}

Attempts to use observations of extragalactic sources for
cosmological studies have shown us that
extreme care is required. All observational biases must be accounted
for and theoretical ideas tested self-consistently, avoiding circular
arguments. This is especially true for GRBs at this stage of our
ignorance about the basic processes involved in their creation,
energizing, particle acceleration and radiation production.
Here we
outline some of the difficulties and how one may address and possibly
overcome them.

Let us assume that there
exists a one-to-one but unknown relation between
$\eiso$ and $\ep$, $\eiso = {\cal E}_0 f(\ep/E_0)$, and that we have a measure
of  $F_{\rm tot}$ and  $z$. Here ${\cal E}_0$ and $E_0$
are some constants, and for convenience we have defined $f(x)$ which is the
inverse of the
function $g$ introduced above. The

From equations  (\ref{eiso}) and (\ref{dm}) we can write
\be\label{omega}
\int^Z_1 dZ'[\Omega (Z')]^{-1/2} =
\left({{f(\epo Z)/E_0)}\over{ZF_{\rm tot}/F_0}}\right)^{1/2}\,\,\,\,\,{
\rm
with}\,\,\,\,\, F_0={{{\cal E}_0}\over{4\pi (c/H_0)^2}}.
\ee
For general equations of
state $P=w_i \rho$, $\Omega(Z)=\sum_{i} \Omega_i Z^{3(1+w_i)}$.
The aim of any cosmological test is to determine
the values of different $\Omega_i$  and their evolutions (\eg changes in $w_i$)
.
If we make the somewhat questionable assumption of complete
absence of cosmological evolutions  of $\eiso, \ep$ and the
function $f(x)$, then this equation involves two unknown
functions $\Omega(z)$ and $f(x)$. {\underline{In principle,}} if the
forms of these functions are known, then one can rely on
some kind of minimum $\chi^2$ method to determine the parameters of
both functions, assuming that there is sufficient data to overcome the
degeneracies inherent in dealing with large number of
parameters.
By now the parametrization of $\Omega(Z)$
has become standard. However,
the form and parameter values of $f(x)$ is
based on poorly understood data and theory,
and currently requires an assumed cosmological
model. Using the {\underline{form}}  ({\it e.g.} the power law used by Gea05)
derived based on an assumed cosmological model to carry out such
a test  is strictly speaking circular. (It is even
more circular to fix the value of  {\underline{parameters}}, in this case the
index $\eta$, obtained in one cosmological model to test others as
done by Dai \ea 2004). Even though different models yield
results with small differences, this does not
justify the  use of circular logic. The differences
sought in the final test using equation \ref{omega} will be of the same order.
The situation is even more difficult because
as stressed above the correlation  is not a
simple one-to-one relation but is a statistical one.
Finally, the most important unknown which plagues all cosmological tests using
discrete sources is the possibility of the existence of an a priory unknown
evolution in one or all of the relevant characteristics. For example, the
intrinsic luminosity $L_{iso}$ might suffer large evolution which we refer to as
luminosity evolution. The value of $E_p$ can also be subject to selection
effects, or the correlation function f(x) may evolve with redshift, i.e.
$\eta=\eta(z)$. For such a general case we are dealing with 4 unknown ??? of the
two above. Moreover the rate function of GRBs most likely is not a constant and
can influence the result s with a broad distribution. We address some of these
questions now.

\section{GRB Evolutions}

For a better understanding of GRBs themselves and the possibility of their use
for cosmological tests we need to know whether characteristics such as $\eiso$,
$\tjet$, $\ep$, the correlation function $f(x)$ and the occurrence rate ${\dot
\rho_{\rm GRB}}$ (number of GRBs per unit co-moving volume and time) change with
time or $Z$. For example to use the $E_p - \eiso$ correlation for cosmological
purposes, one need to first establish the existence of the correlation and
determine its form locally (low redshift).One then has to rely on a theory or
non-circular observations to show that either this relation does not evolve or
if it does how it evolves.The existing GRB data is not sufficient for such a
test. In factthere seems to be some evidence that there is evolution. Lie??? has
shown by subdividing the data into 4 z-bins, he obtained different index $\eta$
which change significantly, rendering previous use of this relation for
cosmological test invalid. This emphasizes the need for a solid understanding of
the evolution of all GRB characteristics. Two of the most important
characteristics are the energy generation $\eiso$ and the rate of GRBs. These
are also two characteristics which can be determined more readily and ??? with
higher uncertainty. In what follows we address these two questions. We will use
all GRBs with known $Z$ irrespective of whether we know the jet angle $\tjet$
because this gives us a larger sample and because in view of new {\it Swift}
observations (Nousek \ea 2006) the determination of the latter does not seem to
be straightforward. Also since it is often easier to determine the peak flux
$f_p$ rather than the fluence threshold, in what follows we will use the peak
bolometric luminosity $L_p=4\pi d_m^2 Z^2 f_p$ instead of $\eiso=\int L(t)dt$.

\subsection{Evolution with Pseudoredshifts}

Before considering GRBs with known redshifts we briefly mention that there has
been two indications of strong evolutionary trends from use of pseudo redshifts
based on the so-called luminosity-variability and lag-luminosity correlations
(Lloyd \ea 2002, Kocevski \& Liang 2006) using the methods developed by Efron \&
Petrosian (1992, 1994). These works show existence of a relatively strong
luminosity evolution $L(z)=L_0 Z^{\alpha}$ ($\alpha= 1.4 \pm 0.5, 1.7 \pm 0.3 $)
from which one can determine a GRB formation rate which also varies with
redshifts and can be compared with other cosmological rates such as the star
formation rate.

\subsection{Evolution with Measured Redshifts}

\subsubsection{Description of the Data}

We have compiled the most complete list of GRBs with known redshift. Since the
launch of the {\it Swift} satellite, this list has become significantly larger.
We include only bursts with good redshift determination meaning that GRBs with
only upper or lower limits on their redshift are not in our sample. 
On total, our sample contains 86 bursts, triggered by 4 different instruments: 
BATSE on board CGRO (7 bursts), BeppoSAX (14), HETE-2 (13), and Swift (52).
For each burst we collected fluence and peak flux in the energy bandpass of the
triggering instrument, as well as the duration of the burst. When available we
have also collected spectral information namely the parameters that define the
Band function; the energy peak ($E_p^{obs}$) as well as the low ($\alpha$) and
high ($\beta$) energy indexes of the $\nu F_{\nu}$ spectrum. When a good
spectral analysis was not available, we took as default values the mean of the
BATSE distributions based on large number of bursts: $<\alpha>=-1.0$,
$<\beta>=-2.3$ and $<E_p^{obs}>=250$ keV.
For non-Swift bursts, all this information was mostly extracted from FB05 data
set to which we added results from recent spectral analysis released in GCN. For
Swift GRBs,
redshift, duration, fluence and peak flux were compiled from the Swift
Information webpage (http://swift.gsfc.nasa.gov/docs/swift/archive) and spectral
information have been retrieved from GCN releases and we have also looked at
spectral analysis ourselves for some of them.
We assumed the following cosmological model:

$$\Omega_M=0.3, \Omega_{\lambda}=0.7, H_0=70 \mbox{ km s$^{-1}$ Mpc$^{-1}$}$$

\noindent
in order to determine the intrinsic properties ({\it e.g.} ${\cal E}_{iso}$ see
eq. [\ref{eiso}]). 
${\cal E}_{iso}$ is here calculated for a rest-frame bandpass [20,2000] keV.
Note that K-correction due to the shift of the photons into the instrument
bandpass has been properly taken into account for the ${\cal E}_{iso}$
calculation (see Bloom \ea 2001). From this, we calculate the average
isotropic-equivalent luminosity as:

$$L_{iso}=\frac{\eiso}{T_{90}}$$

\noindent
when $T_{90}$ is the duration of the burst that includes 90$\%$ of the total
counts.

\begin{figure}[htbp]
\begin{center}
\includegraphics[width=14cm]{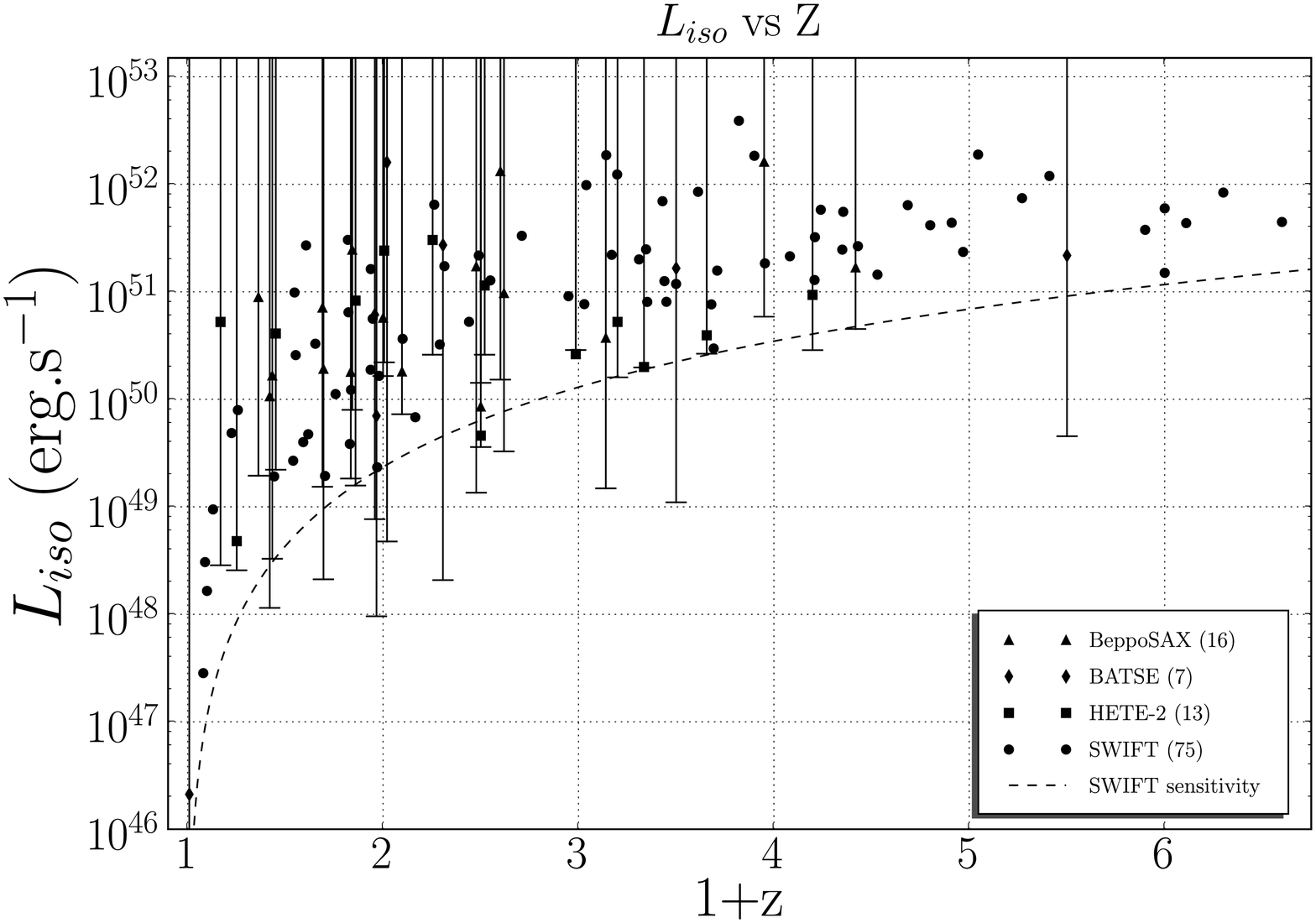}
\end{center}
\caption{Isotropic average luminosity versus redshift for all bursts in our
sample (86). Different symbols represent bursts observed by different
instruments: BATSE(7), Beppo-SAX (14), HETE-2 (13), {\it Swift} (52). For all
non-{\it Swift} burst, a vertical line is plotted representing the range of
isotropic luminosity in which the burst would still have been observable by the
instrument keeping all its others parameters fixed. Using the work of Lamb \&
al. 2005, the limiting luminosity is taken to be only dependent on the energy
peak $E_p$ of the bursts.  For {\it Swift} bursts, a conservative threshold flux
of 0.8 ergs s$^{-1}$ cm$^{-2}$ has been chosen. This limit is shown as a dashed
line.}
\label{3}
\end{figure}

Because different instruments have been used to collect this information, the
sample is very heterogeneous and suffers from various selection and truncation
effects that vary from burst to burst. The most simple of these truncation
effects is due to the limiting sensitivity of the instruments. A GRB trigger
will occur when the peak flux of the burst exceeds the average background
variation by a few sigmas (depending on the setting of the instrument).
In an attempt to carefully take into account this effect into our study, we used
the analysis carried out by Lamb \ea 2005 for pre-{\it Swift} instruments. In
this analysis, they computed the sensitivity for each instrument depending on
the spectral parameters of the bursts. Therefore, for each specific burst of our
data set, from its spectral parameters it is possible to determine the limiting
photon flux of the instrument (for specific GRB with its specific spectral
parameters). From these, we can easily compute the limiting peak flux
$f_{p,lim}$ for our burst (assuming the Band function for our spectrum).
Finally, we can determine the detection threshold of the observed energy fluence
$F_{obs}$. This lower limit $F_{obs,lim}$ is obtained via the simple
relationship (Lee \& Petrosian 1996):

$$ \frac{F_{obs}}{F_{obs,lim}} =  \frac{f_{p}}{f_{p,lim}} $$

\noindent
Using the same reasoning we can obtain the limiting values for the intrinsic
quantities meaning the intrinsic values that a given burst needs to have in
order to be detected:

$$ \frac{\eiso}{{\cal E}_{iso,lim}} =\frac{L_{iso}}{L_{iso,lim}} = 
\frac{f_{p}}{f_{p,lim}} $$

\noindent
Those limiting average luminosities for each bursts of our sample are
represented in Figure \ref{3}. This analysis was not carried out for BAT
instrument on board Swift therefore we used a conservative threshold of $0.8$ erg
s$^{-1}$ cm$^{-2}$ for all of Swift bursts.

\subsubsection{Analysis and results}

\begin{figure}[htbp]
\begin{center}
\includegraphics[width=14cm]{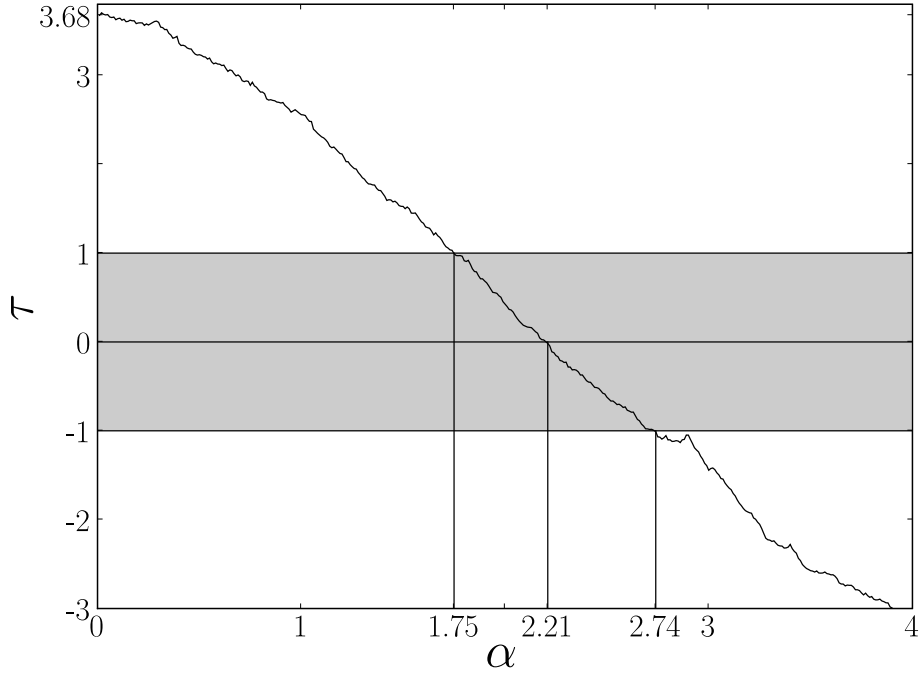}
\end{center}
\caption{Variation of the $\tau$ parameter with the power law index $\alpha$ of
the luminosity evolution. $\tau=0$ means no correlation which gives the best
value of $\alpha = 2.21$ for the assumed power law form with a one sigma range
of $1.75$ to $2.74$.}
\label{5}
\end{figure}

We now describe our determination of luminosity and density rate evolution of
the parent population of our GRB sample. Our analysis is based on the work done
by Efron \& Petrosian (1992, 1994). We refer the reader to these two papers for
details. We will here simply describe the most important steps of the analysis
and what it allows us to infer on our data sample. 
This method has been developed in order to take into account effects of data
truncation and selection bias on a heterogeneous sample from different
instruments with different sensitivities as described above and shown in Figure
\ref{3}. The method corrects for this bias by applying a proper rankings to
different subset of our sample.
The first step is to compute the degree of correlation between the isotropic
luminosity and redshift. For that we use the specialized version of Kendell's
$\tau$ statistics. The parameter $\tau$ represents the degree of correlation
found for the entire sample with proper accounting for the data truncation.
$\tau=0$ means no correlation is found between the two parameters being
inspected (luminosity and redshift in our case). Any other specific value
$\tau_0$ implies presence of a correlation with a significance of $\tau_0 
\sigma$. 
With this statistic method in place, we can calculate the parametrization that
best describe the luminosity evolution. To establish a functional form of the
luminosity-redshift correlation, we assume a power law luminosity evolution:
$L(z)=L_0 Z^{\alpha}$.  We then remove this dependency from the observed
luminosity: $L'  \rightarrow L_{observed}/(1+z)^{\alpha}$ and calculate the
Kendell's $\tau$ statistics as a function of $\alpha$. Figure \ref{5} shows the
variation of $\tau$ with $\alpha$.

Once the parametric form for the luminosity evolution have been determined, this
nonparametric maximum likelihood techniques can be used to determine the
cumulative distribution for luminosity and redshift (see Efron \& Petrosian
1994) say $\Phi(L)$ and $\sigma(z)$, which gives the relative number of bursts
under a certain redshift z. From this last function, we can easily draw the
comoving rate density $\dot{n}(z)$, which is the number of GRB per unit comoving
volume and unit time: 	

\be
\label{prout1}
\dot{n}(Z)=\frac{d \sigma(Z)}{dZ} \frac{Z}{dV/dZ}
\ee

where the factor $Z$ is to take the time dilatation into account. Note that this
method do not provide any constrain on the normalization of any of the quantity
mentioned above. Normalization will therefore be set arbitrarily on all our
figures representing these functions.

We find a 3.68 $\sigma$ evidence for luminosity evolution (see the $\tau$ value
at the onset of Figure \ref{5} when $\alpha=0$). From this figure, we can also
infer that $\alpha = 2.21$ for value obtained when $\tau = 0$ gives the best
description of the luminosity evolution for the assumed form has a one sigma
range of $[1.75,2.74]$. Constrain on the $\alpha$ parameter is not very tight
off course since the size of our sample is still limited. While current
satellites accumulate more data, we will be able to increase our data set and
further constrain this parameter in the future.

\begin{figure}[htbp]
\begin{center}
\plottwo{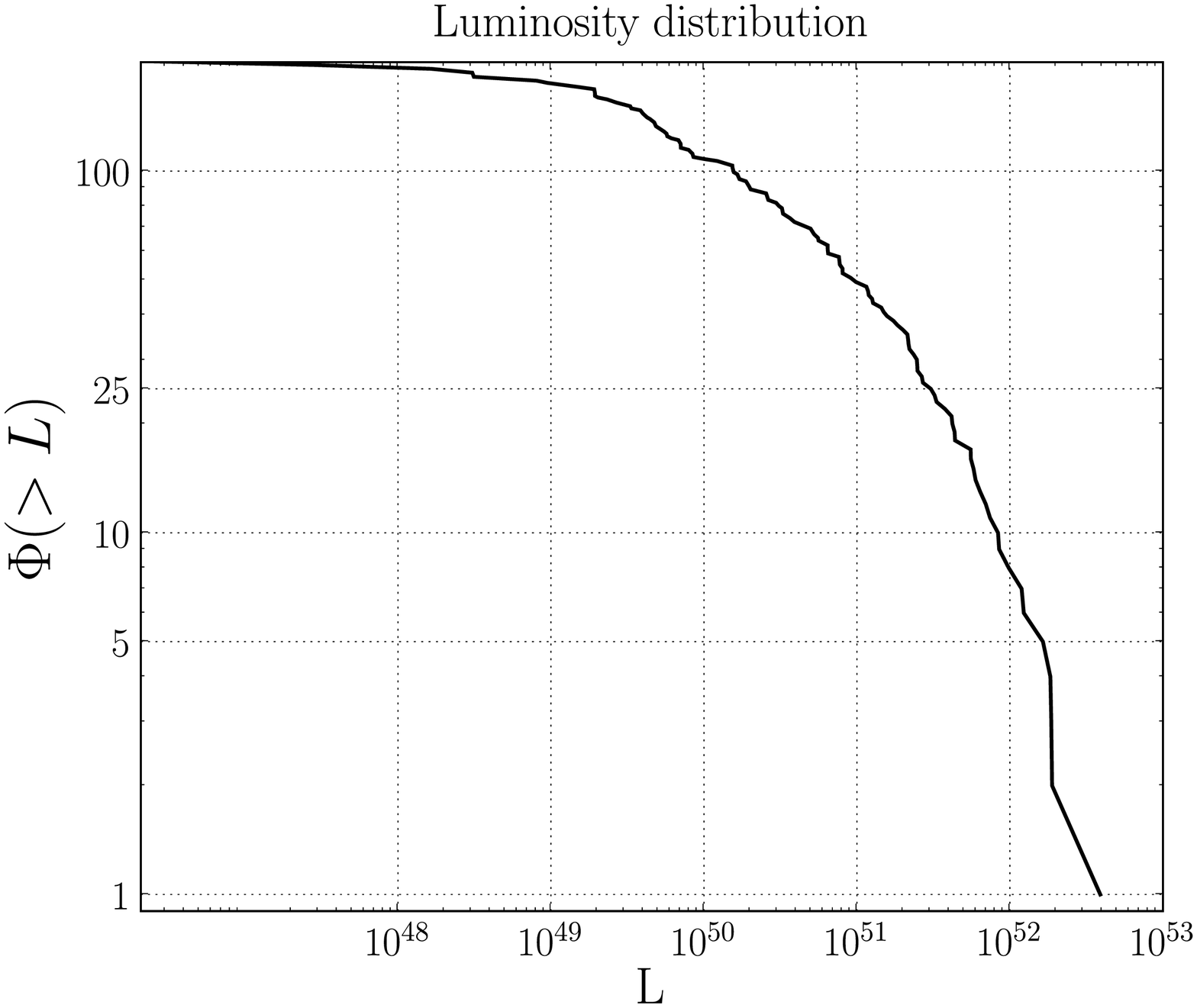}{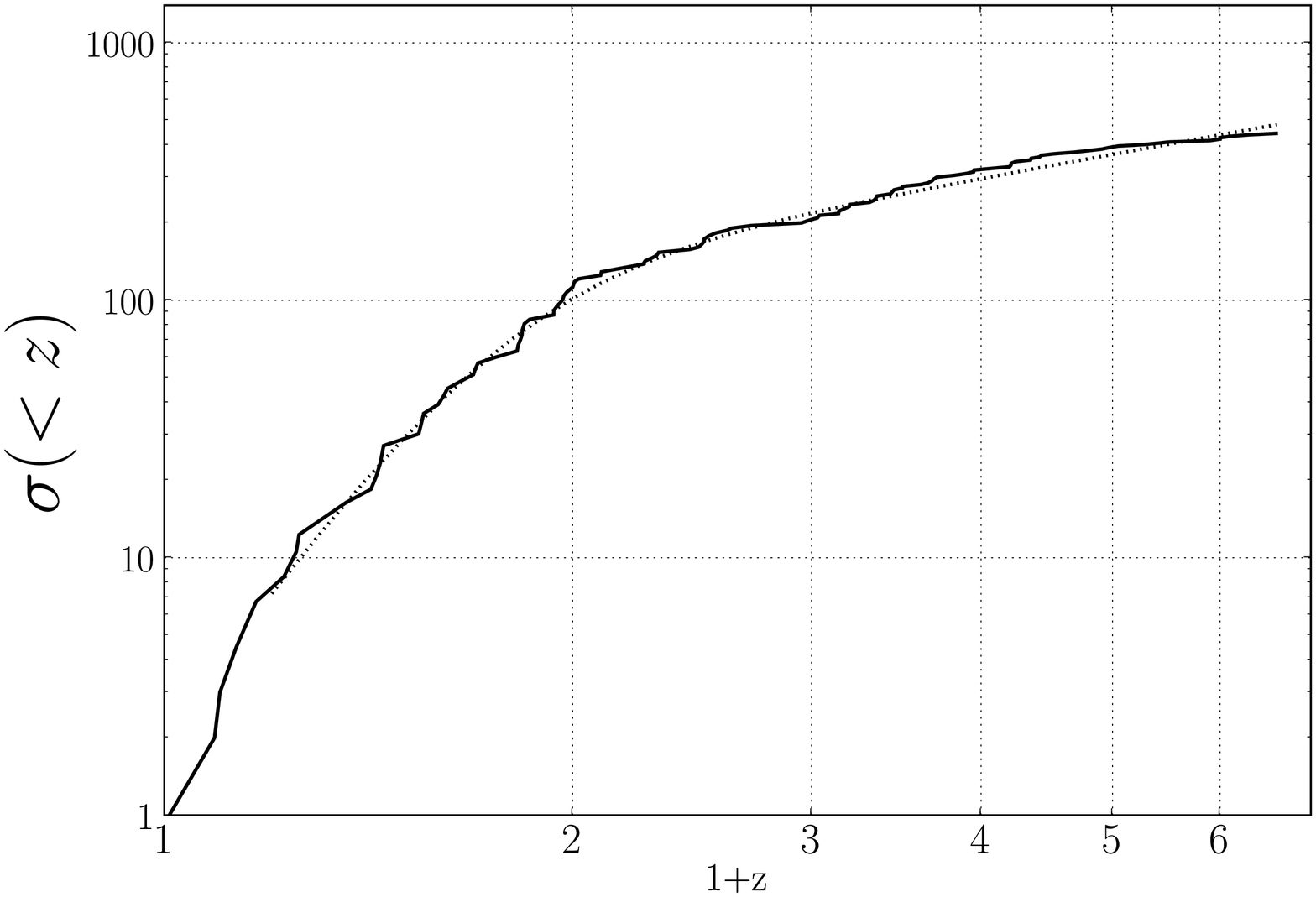}
\end{center}
\caption{The cumulative luminosity distribution $\Phi(L)$ (left panel) and the
cumulative redshift distribution $\sigma(z)$ (right panel). Fit to the cumulative
density rate is represented with a point line.}
\label{4}
\end{figure}

\begin{figure}[htbp]
\begin{center}
\includegraphics[width=14cm]{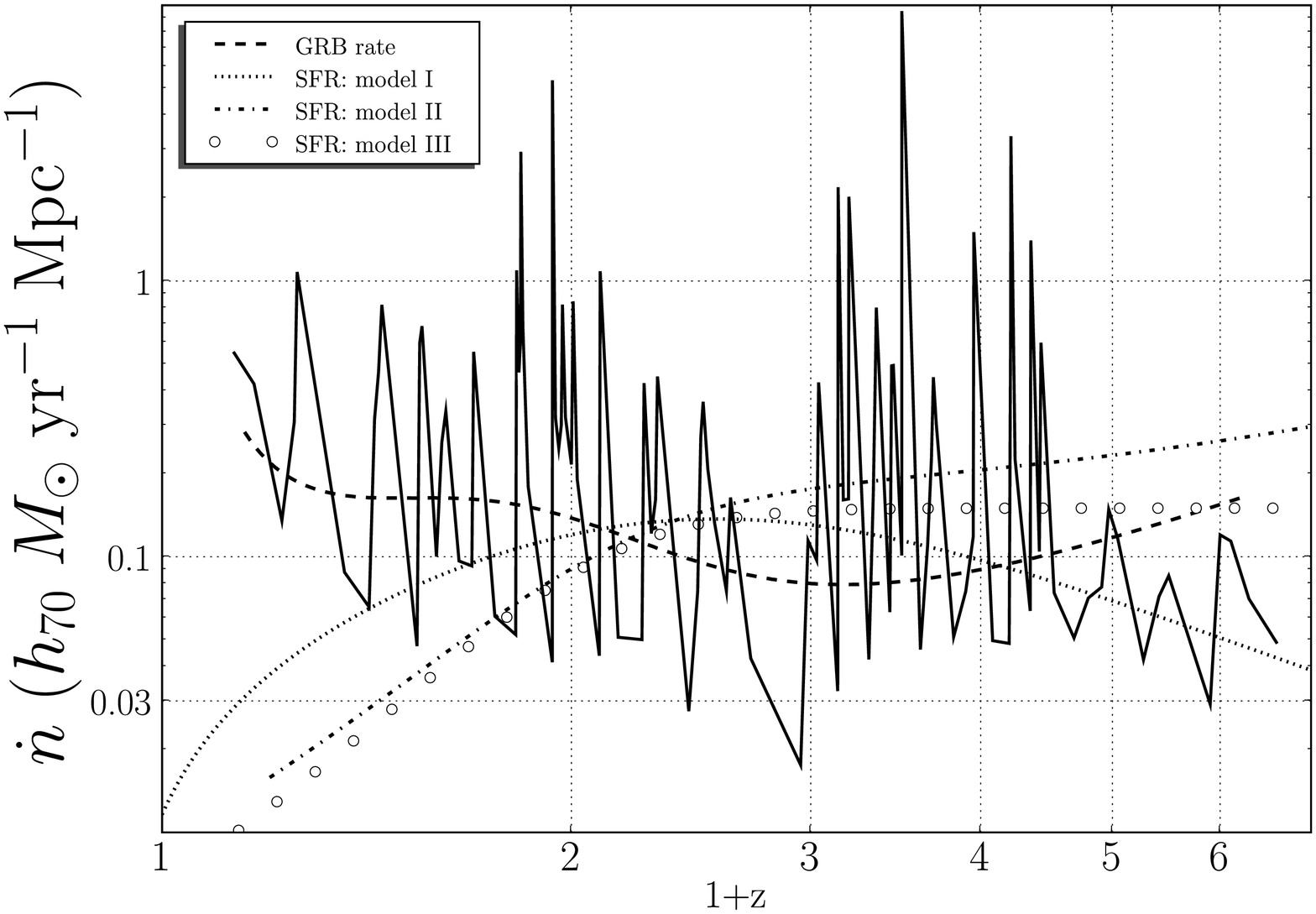}
\end{center}
\caption{The comoving rate density $\dot{n}(z)$. The dashed line was obtained by
fitting the cumulative density distribution by the parametrized smooth function
of equation \ref{fitting}. We also show comparison of the density rate (from
Figure \ref{6}) with three different SFR scenarios taken from literature. No SFR
scenario seems to match the density rate deduced from our analysis.}
\label{6}
\end{figure}

\begin{figure}[htbp]
\begin{center}
\includegraphics[width=14cm]{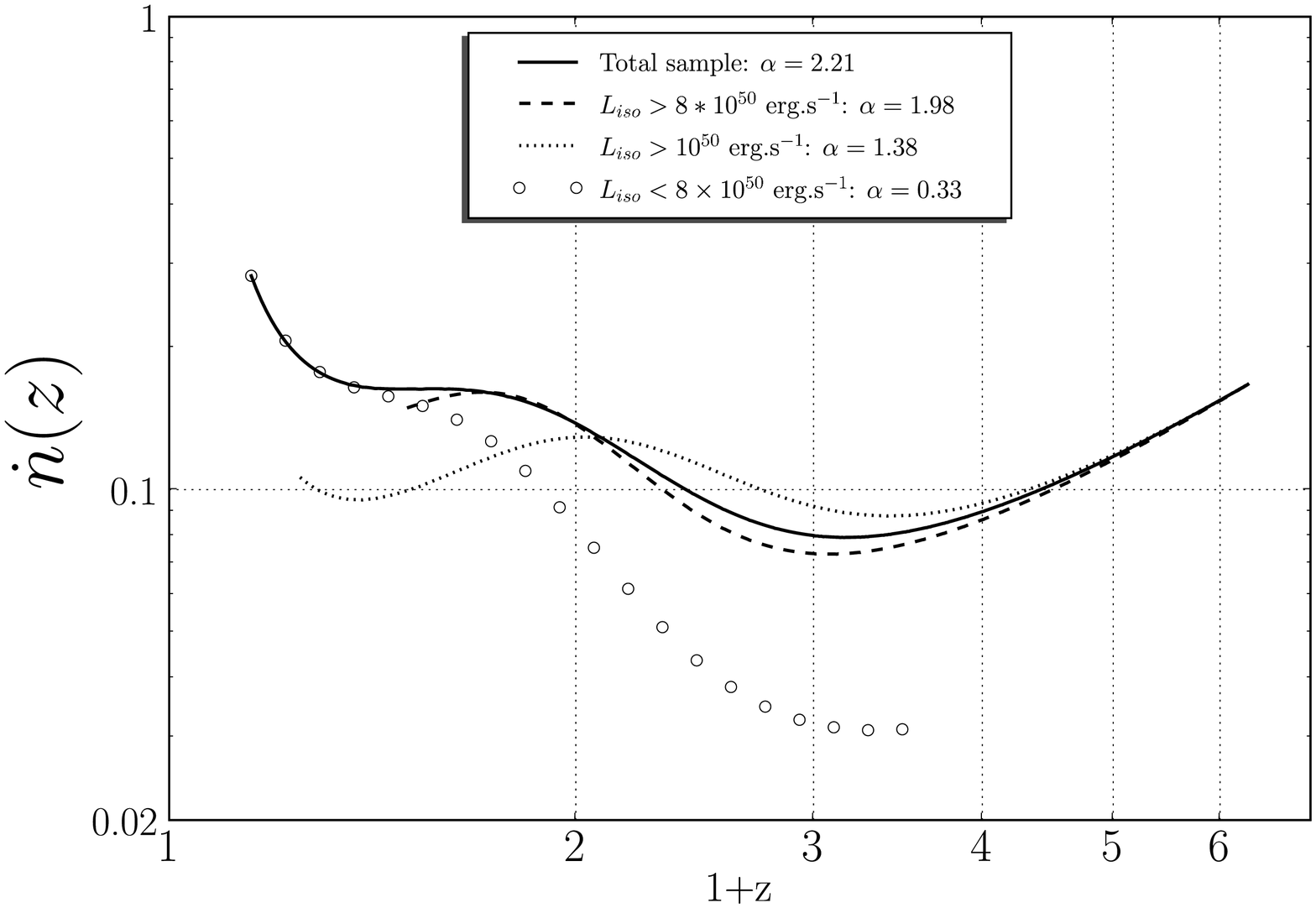}
\end{center}
\caption{Comparison of the comoving density rate evolution of the total sample
with that of several sub-sample where we impose three different luminosity
thresholds: $L_{iso}>10^{49}$, $>10^{50}$, and $>8 \times 10^{50}$
ergs.s$^{-1}$. We also looked at a low luminosity population where we imposed of
maximal luminosity of $8 \times 10^{50}$ ergs.s$^{-1}$. Each sub-sample is
subject to the same analysis and has provided different luminosity evolution as
evident from the different values of $\alpha$. As expected, the rate at low
redshift decreases with increasing values of threshold.}
\label{7}
\end{figure}

The cumulative functions are both shown in Figure \ref{4} and the estimated
comoving density rate is shown by the jagged curve in Figure 6. Most of the high
frequency variation is not real and is due to taking the derivative of a noisy
curve ($\sigma (Z)$). The dashed line was obtained by fitting the cumulative
density distribution by the following parametrized function:

\be
\label{fitting}
\sigma(Z) \propto \frac{(Z/Z_0)^{p_1}}{(1+Z/Z_0)^{p_1-p_2}}
\ee

\noindent
with the following values for the parameters: $Z_0=1.8$, $p_1=7.1$, and
$p_2=0.95$

These results are still very preliminary as more data become available, accuracy
of the density function will increase constraining further the evolution rate of
long bursts. By tackling this problem for the first time we hope to set the
ground for further analysis in the future.

The behavior of the comoving density rate for our sample of long bursts is quite
peculiar with a significant rate increase happening at low redshifts. This
effect might be due to some selection effects that we have not included in our
analysis. For instance, it might be a consequence of the fact that instruments
detect more easily low-redshifts host galaxies and therefore create a bias
toward low redshifts GRBs. Another interesting feature is the steady increase we
obtain in the GRB rate at high redshifts ($z > 3$). Figure 7 compares the
estimated comoving rate evolution with different models of Star Formation Rates
(SFRs I, II, III).

For comparison with Star Formation history, we used three different models taken
from the literature:

- Steidel et al. 1999:

\be
\dot{n}=0.16 h_{70} \frac{e^{3.4 z}}{e^{3.4 z} +22} M_{\odot} \mbox{yr}^{-1}
\mbox{Mpc}^{-3}
\ee

- Porciani \& Madau 2000:

\be
\dot{n}= 0.22 h_{70} \frac{e^{3.05 z-0.4}}{e^{2.93 z}+15} M_{\odot}
\mbox{yr}^{-1} \mbox{Mpc}^{-3}
\ee

- Cole \ea 2001:

\be
\dot{n}= \frac{(a + b z) h_{70}}{1+(z/c)^d} M_{\odot} \mbox{yr}^{-1}
\mbox{Mpc}^{-3}
\ee

with $(a, b, c, d) = (0.0166, 0.1848, 1.9474, 2.6316)$

As evident, no SFR scenario seems to match the density rate evolution deduced
from our analysis, specially at low redshift. How much of this difference is
real and how much is due to other selection effects that we have not quantified
is unclear. Because of increasing difficulty of identifying the host galaxy with
increasing redshift one would expect some bias against detection of high
redshift bursts. But the largest densities for SFR seems to be in the
intermediate redshift range.

An other possibility is that there may exist subclasses of GRBs such as low or
high luminosity classes. In order to test this eventuality we have defined
several subsets of our total GRB sample carried out the above analysis for each
subsamples, determining a new luminosity evolution (a new $\alpha$) and then
proceeding to obtain $\dot{n} (z)$ from the smooth function fitting $\sigma
(z)$. We impose different luminosity thresholds for the different subsamples.
Three different threshold have been chosen: $L_{iso}>10^{49}$ ergs.s$^{-1}$,
$L_{iso}>10^{50}$ ergs.s$^{-1}$, and $L_{iso}>8 \times 10^{50}$ ergs.s$^{-1}$.
We also looked at a low luminosity population where we imposed of maximal
luminosity of $10^{50}$ ergs.s$^{-1}$. Figures \ref{6} and \ref{7} compare the
new rates with that of the total sample. As expected high luminosity samples
contribute less to the rate at low redshifts. But the general trend and the
differences with SFR are essentially still present. However the method and
framework we presented would be a very valuable tool when enough data has been
accumulated.

\section{Summary and Conclusion}

We have considered GRBs as cosmological tools. We find that GRBs are not SCs and
the correlations found so far are statistical in nature and too broad to be very
useful for cosmological model parameter determination. In addition we have shown
that there is strong evidence for evolution of the peak luminosity of GRBs. This
indicate a very likely possibility that $\eiso$ and $\egamma$ may also have
undergone comparable changes. There may also be evolution of $\ep$, $\tjet$ or
other relevant characteristics. This makes use of GRBs as cosmological tools
more difficult.

We may therefore ask is this process hopeless? Strictly speaking the answer is
no. Some
broad brush conclusion can already be reached.
For example, one can test the relative merits of different forms for $f(x)$.
As shown by FB05 the SC assumption
($\phi(\eiso)\rightarrow\delta(\eiso - {\cal E}_0)$)
gives unacceptable fit to essentially all cosmological models, but the
use of the power-law form agrees with  Lemaitre type model (FB05, Gea05)
with a relatively long quasi-static
phase (refereed to as a loitering model). Such models, which were in vogue some
time
ago (see Petrosian 1974), are currently unacceptable because of their
low (baryonic plus dark) matter  density and large curvature.
This indicates that the form of the correlation and/or other assumptions ({\it
e.g.} no
evolution) are not correct. In this paper, we have shown that there is strong
evolution of luminosity and $\eiso$. Lei \ea have also shown that the form of
the $E_p-\eiso$ correlation may evolve. These are tentative results and more
data are required to determine  these evolution trend and their meaning.

The relevant variables in addition to redshift are $\ep^{obs}=\ep/(1+z)$ and
$F_{tot}$, determination of which requires a good description of the
total spectrum. We need to
know the observational selection biases for all the variables and
their parameters, and use accurate statistical methods to account for the
biases and data truncations. From these one can learn about the distributions of
$\eiso$ and $\ep$ and their correlations.

Obviously {\it Swift} observations will be extremely helpful
and eventually may provide the data required for this complex
task. In the near future, however, from analysis of the incoming and
archived data we will (and need to) first learn more about the nature of the
GRBs than
cosmological models.  Eventually we may have
enough information to construct a well defined ``SC", which can be used for
global cosmological tests as is
done using type Ia supernovae.   The immediate situation may be more analogous
to
galaxies where the cosmological tests
are rendered complicated because of the multivariate situation and broad
distributions of the relevant variables.
Consequently, over the years the
focus of activity has shifted
from the determination of the few global cosmological parameters to the
investigation of structure formation, the building
process of the black holes  and the star formation  rate (SFR). Similarly, we
expect that
from investigation of GRBs we will learn about the evolution, distributions and
correlations of their intrinsic characteristics, and the
relationship of these with the evolutionary rates of other
cosmological sources and the formation rates of stars, supernovae and black
holes.

In summary, on the long run cosmological test with
GRBs may be possible, either carried out with clever statistical
methods, or by identification of  a subclass of ``SCs''. On
a shorter time scale, we need to learn more about the intrinsic
characteristics of GRBs, and provide a reasonable
theoretical interpretation for them and their cosmological evolution. As an
example, as shown here, one outcome of our analysis is the determination of the
evolution rate of GRBs. We have shown that for all GRBs with known redshifts
this rate appears to be different form the SFR. These differences seems to be
present with different subdivision of the sample and may be consequences of
other selection biases not included in our analysis. But the difference is not
what one would expect from some possible observational selection...??

\section{REFERENCES}
\baselineskip=0pt
\renewcommand{\bibitem}{\noindent\hangindent=2pc \hangafter=1}

\bibitem{}Amati, L. \ea 2002, A\&A 390, 81

\bibitem{}Atteia J.-L. \ea 2004, AIP Conf. Proc. 727, eds,
  E.E. Fenimore \& M. Galassi, p 37

\bibitem{}Band, D. L. \& Preece, R. D. 2005, (astro-ph/0501559)

\bibitem{}Cen, R., Fang, T. 2007, (astro-ph/0710.4370)

\bibitem{}Cole, S. 2001, MNRAS, 326, 255

\bibitem{}Dai, Z. G., Liang, E. W. \& Xu, D. 2004, ApJ, 612, L101

\bibitem{}Efron, B.  \& Petrosian, V.  1992, ApJ, 399, 345

\bibitem{}Efron, B.  \& Petrosian, V.  1994, ApJ, 399, 345

\bibitem{}Efron, B.  \& Petrosian, V.  1999, J. Am. Stat. Assoc., 94, 824

\bibitem{}Fenimore, E. E. \& Ramirez-Ruiz, E. 2000 (astro-ph/0004176)

\bibitem{}Frail, D. A. \ea 2001, ApJ Letters, 562, L65

\bibitem{}Friedman, A. S. \& Bloom, J.S. 2005, ApJ, 627, 1;  {\bf FB05}

\bibitem{}Ghirlanda, G., Ghisellini, G. \& Lazzeti, D. 2004a ApJ, 616,
  331

\bibitem{}Ghirlanda, G. \ea 2004b, ApJ, 613, L13
  331

\bibitem{}Ghirlanda, G. \ea 2005a, (astro-ph/0504306):  {\bf GGFLA05}

\bibitem{}Ghirlanda, G., Ghisellini, G. \& Firmani, C. 2005b, (astro-ph/0502186)

\bibitem{}Kocevski, D. \& Liang, Edison 2006; astro-ph/0601146

\bibitem{}Lamb, D.Q. \ea 2004, New Ast. Review, 48, 459

\bibitem{}Lamb, D.Q., Donaghy, T.Q. \& Graziani, C.  2005, ApJ, 620, 355

\bibitem{}Lee, T.T. \& Petrosian, V. 1996, ApJ, 470, 479

\bibitem{}Lee, T.T. \& Petrosian, V. 1997, ApJ, 474, L37

\bibitem{}Lloyd, N. M., Petrosian, V. \& Mallozzi, R.S. 2000, ApJ, 534,227: {\bf
LPM00}

\bibitem{}Lloyd, N. M. \& Petrosian, V. 1999, ApJ, 511, 550 {\bf LP99}

\bibitem{}Lloyd, N. M., Fryer, C. L. \& Ramirez-Ruiz, E. 2002, ApJ,  574,554

\bibitem{}Mallozzi, R. \ea 1995, ApJ, 454, 597

\bibitem{}Nakar, E. \& Piran, T. 2004 (astro-ph/0412232), 2005,
(astro-ph/0503517)

\bibitem{}Norris, J. P., Marani, G.F. \& Bonnell, J.T. 2000, ApJ, 534, 248

\bibitem{}Norris, J. P. 2002, ApJ, 579, 386

\bibitem{}Nousek, J. 2006, Paper presenterd at the ``Challenges of Relativistic
Jets", Krakow, Poland.

\bibitem{}Petrosian, V. 1974  Proc. of IAU Symp. 63, ed. M.S. Longair, p. 31-46.

\bibitem{}Petrosian, V. 1998, ApJ, 507, 1

\bibitem{}Petrosian, V., Bouvier, A. \& Ryde, F. 2006; paper presented at COSPAR
in Beijing, July 2006

\bibitem{}Porciani, C. , Madau, P. 2001, ApJ, 548, 522

\bibitem{}Reichart, D. E. \ea 2001, ApJ, 552, 57

\bibitem{}Steidel \ea 1999, ApJ, 519, 1

\end{document}